# Performance Comparison of Persistence Frameworks

Sabu M. Thampi[*]
*Asst. Prof., Department of CSE*
*L.B.S College of Engineering*
*Kasaragod-671542*
*Kerala, India*
*smtlbs@yahoo.co.in*

Ashwin A.K
*S8, Department of CSE*
*L.B.S College of Engineering*
*Kasaragod-671542*
*Kerala, India*
*ashwin_a_k@yahoo.co.in*

## Abstract

*One of the essential and most complex components in the software development process is the database. The complexity increases when the "orientation" of the interacting components differs. A persistence framework moves the program data in its most natural form to and from a permanent data store, the database. Thus a persistence framework manages the database and the mapping between the database and the objects. This paper compares the performance of two persistence frameworks – Hibernate and iBatis's SQLMaps using a banking database. The performance of both of these tools in single and multi-user environments are evaluated.*

## 1. Introduction

When a component based on one kind of approach (e.g. object oriented) tries to interact directly with another object having its roots in another kind of approach (e.g. relational), the complexity increases due to the knots and knaves of cross approach communication. This is evident in all the database APIs provided by different languages. The best example of this is the Java Database Connectivity (JDBC) API. Though JDBC provides an easy method for accessing different databases without much ado, it is basically a low level API providing only a thin layer of abstraction. This is adequate for small and medium projects, but is not well suited for enterprise level applications. With JDBC opening and closing the connection involves a lot of code. What is required is a framework that can act as a mediator between both parties.

In OOP, it is typically the behavior of objects (usecases, algorithmic logic) being emphasized. On the other hand, it is the data that counts in database technology. This fact serves as a common motive for the combination of these two paradigms [1]. The core component of this coupling is what is called "object-relational mapping" which takes care of the transitions of data and associations from one paradigm into the other (and vice versa). In order to make a program's object persistent, which means to save its current state and to be able to load that data later on, it is necessary to literally map its attributes and relations to a set of relational tuples. The rules defining such mappings can be quite complex. Here, the term "mapping" can be defined as the application of rules to transfer object data to a unique equivalent in an RDBMS (relational database management system) and vice-versa. Viewed from the object's perspective, this ensures that all relevant object data can be saved to a database and retrieved again.

A persistence framework moves the program data in its most natural form (in memory objects) to and from a permanent data store, the database. The persistence framework manages the database and the mapping between the database and the objects. Persistence framework simplifies the development process. There are many persistence frameworks (both Open Source and Commercial) in the market. Hibernate and iBatis are examples for ORM frameworks for Java.

Hibernate [2] is an open source project being covered by BossTM. It is intended to be a full-scale ORM environment and features interesting functionality, such as "real transparency": a data class does not have to extend special classes of Hibernate; it only has to make properties available through standard get-/set-methods. Hibernate uses bytecode processing to extend from these classes and implement persistence. It also supports – according to the project homepage – a sophisticated caching mechanism (dual-layer, which can be distributed as well) using pluggable cache providers. Hibernate is an object/relational persistence and query service for Java. Hibernate lets you develop persistent classes following common Java features - including association, inheritance, polymorphism, composition and the Java collections framework. The Hibernate

---
[*] Corresponding Author

Query Language, designed as a "minimal" object-oriented extension to SQL, provides a bridge between the object and relational worlds. Hibernate also allows you to express queries using native SQL or Java-based Criteria and Example queries.

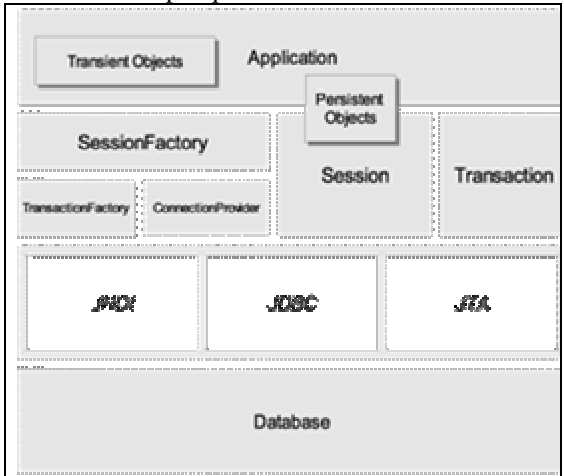

**Figure 1: Full Cream Architecture of Hibernate**

The SQLMaps product of iBatis [3,4] does not represent an ORM environment at the scale of Hibernate. Just as the name suggests, it is heavily SQL-centric and provides means to access centrally stored SQL-statements in a convenient way. The mapping functionality is able to create objects based on query data, but there is no transaction support. The SQLMaps product is light-weight and is expected to run faster than the heavy loaded full-scale ORM toolsIt uses a special mapping files in which the developer should expose object's properties to be made persistent as well as respective database tables and columns these properties should be mapped to. In addition to that there are something called dynamic queries, caching of queries, transactions and calling stored procedures. The framework maps JavaBeans to SQL statements using a XML descriptor.

The performance comparison of Hibernate and iBATIS are explored in this paper. Both of the above tools have their advantages and disadvantages.

The remaining sections of the paper are organized as follows. Section 2 gives an overview of protype banking application. Simulation results are presented in section 3. Section 4 concludes the paper.

## 2. Online Banking system

A very simple prototype version of an online banking application using struts framework and iBatis/ Hibernate framework is developed to analyze the performance of persistence frameworks. The Jakarta Project's Struts framework, version 1.1b2, from Apache Software Organization is an open source framework for building web applications that integrate with standard technologies, such as Java Servlets, JavaBeans, and JavaServer Pages. Struts offer many benefits to the web application developer, including Model 2 implementation of Model-View-Controller (MVC) design patterns in JSP web applications. The MVC Model 2 paradigm applied to web applications separate display code (for example, HTML and tag libraries) from flow control logic (action classes).

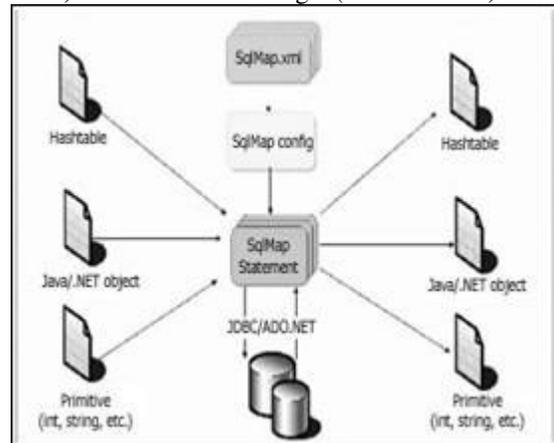

**Figure 2: iBATIS Data Mapper framework**

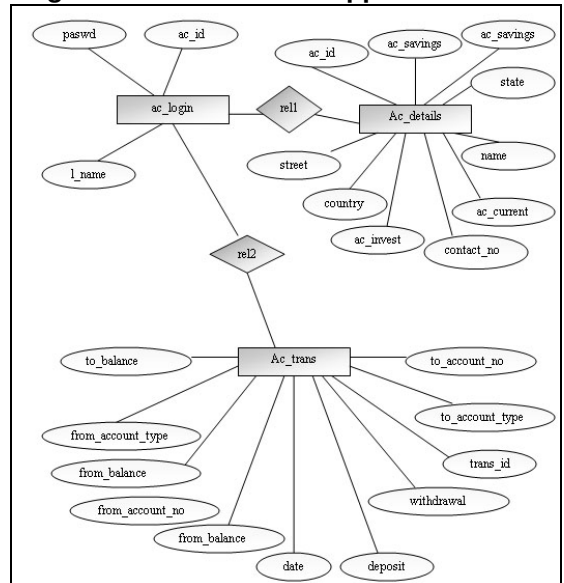

**Figure 3: E-R Diagram for banking application**

The Banking application has a number of functionality such as summary, account details, transfer, transactions, update and contact details. The entity relationship diagram in figure 3 illustrates the relationship among different entities in the prototype system.

The figure 4 shows the architecture of the Demo Banking application. The request is given through the browser. The DAO layer contains only method names.

When the username and password are entered through GUI and the login button is clicked, the application through the struts framework calls for a login action class which first generates a hash code with the entered data. This is supposed to be the account id. Then the application checks the validity of the entered data by querying from the database based on account id. If they are correct, entry is given otherwise an error page is displayed. The jsp page stores the value of the account id to be referenced in future pages.

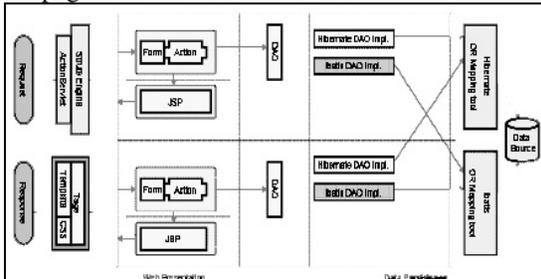

**Figure 4: Architecture of Banking application**

**Figure 5: Account Summary Page**

Each account type is a link. When a particular link is clicked the account id and the account number corresponding to the particular link is passed to transactionaction class. From here the transaction list corresponding to the respective account number is retrieved and is passed to the jsp. There is a unique id for each transaction, which is denoted by Transid.

When the user clicks the transfer button in the jsp page the *pretransferaction class* is called which checks the account types belonging to the account holder from the data table named Ac_details and displays it in a drop down menu. The holder can transfer money to others account as well as to the different account types he has.

## 3. Simulation Results

Performance between Hibernate and iBatis is measured using a java program which uses both hibernate and iBatis to perform basic sql operations on the banking database and the RTT (Round Trip Time) is calculated and used to measure the way these mapping tools perform under various situations. The aim is to get the time from generation of sql to querying bank database and then getting back the data. The program was run from one system and the SQL Server was located in another system. The conditions were the same for both the Hibernate and iBatis. The test also included simulation for a single user and multi user. The simulation of multi user was made through the creation of threads. Java supports multi-threading environment. The number of threads is passed as input to the program. The response of Hibernate and iBatis under multi user environment is monitored. The RTT is monitored for both the cases.

**Figure 6: An example of Transaction details of Banking Application**

The tests were conducted in the following environment:

Operating system: Microsoft Windows 2000
Processor: Intel Xeon 4 Processor
Memory: 1024 MB DDR RAM

The following inputs are needed for the test program:
- **Whether hibernate or iBatis:** The user can specify which DAO to be executed whether Hibernate or iBatis.
- **Number of records:** The number of records to be inserted, deleted, updated is also given at a time.
- **Insert, update, select, delete or all the operations specified:** The user can also specify what operation to be monitored whether insert, update, delete or all operations together can be done.

- **Number of iterations:** Number of times the particular set is to be repeated can also be given as input.
- **Number of threads:** This simulates number of user accessing the application.

The data shown below is how the raw data is recorded and stored into a text file. It is this data that is summarized into graph in figures 8.

Average time (with 5000 records, 10 iterations & 50 threads for hibernate)
    Avg_Insert=3917
    Avg_Update=1462
    Avg_Select(First Time)=37182
    Avg_Select(Second Time)=2361
    Avg_Delete=1414
--------------------------------------------
Average Time (With 5000 Records, 10 Iterations & 50 Threads For Ibatis)
    Avg_Insert=6272
    Avg_Update=5556
    Avg_Select(First Time)=5197
    Avg_Select(Second Time)=5157
    Avg_Delete=5414
--------------------------------------------

Avg_Insert,Avg_Update,,Avg_Select (First Time), Avg_Select (Second Time),Avg_Delete corresponds to average time taken for insert, update, select1, select2 & delete.

The above values are computed as follows:
i. Find the time taken for an operation in each set of record is noted.
ii. The sum of the time taken for all the iterations is found.
iii. The average for that set of iterations is computed.
iv. If multiple threads (say x no of threads) are present we will have many number of averages (here x)
v. The final values are obtained by computing the averages of all the averages previously obtained in step iv.

In the figure 7, y-axis represents time in milli seconds and the x-axis represents the various operations performed (such as insert, update, select1, select2) for both hibernate and iBatis. The graph shows that there are minute differences the time taken between hibernate and iBatis except for select1. The large variation in time taken for select1 is caused due to the complex caching algorithms employed by hibernates. Such techniques have proved to be useful in case of subsequent searches as seen in the graph.

The graph in figure 8 shows the results of time taken when there are 5 threads 5000 records and 10 iterations. As shown in the graph the time taken for hibernate for the first select is very large compare to iBatis, but in all other cases hibernate has an upper hand over iBatis in terms of time taken. Even in the second select operation time taken by hibernate is less compared to that of iBatis. This implies that barring the initial overhead caused by hibernate during the first select it fares well compared to iBatis.

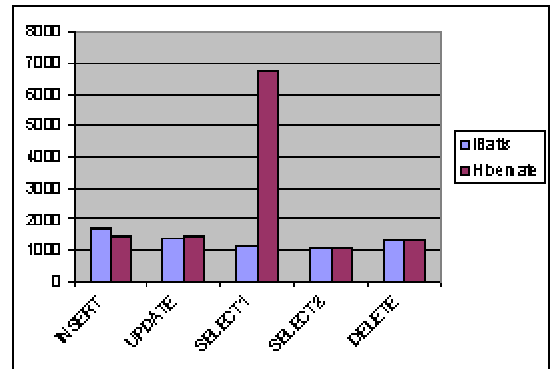

**Figure 7: 1 Thread 5000 records 10 iterations**

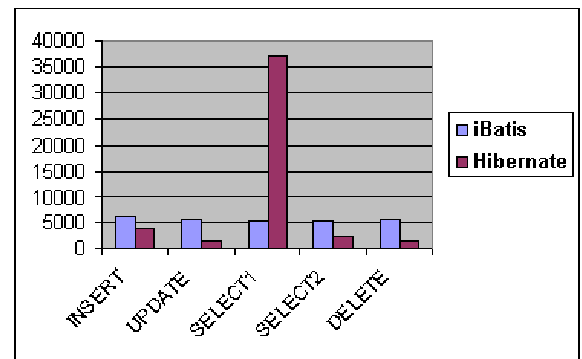

**Figure 8: 5 Threads 5000 records 10 iterations**

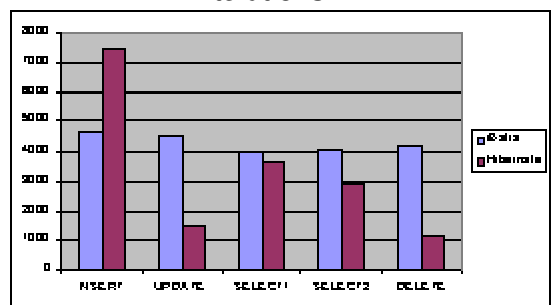

**Figure 9: 50 Threads 5000 records 10 iterations**

Figure 9 represents the time taken when there are 50 threads involved for 5000 records and 10 iterations. In this case it is seen that the large variation that was noticed in case of select1 operation in figures 7 and 8 has now been minimized. When the number of threads was increased to simulate multiple user environments, it is seen that in this case it is iBatis, which lags behind, hibernate compared to the previous cases. The only operation in which hibernate consumes more time is for the insert operation.

## 4. Conclusion

Object relational mapping became important due to increasing coupling between relational database management systems and object oriented application concepts and development. There are tools to automate these mapping tasks, which can be distinguished by the degree to which they abstract the storage logic for the application. Choosing a suitable product can significantly cut down development efforts, costs and time. After conducting the DAO tests on banking database and comparing a similar application using hibernate and iBatis we come to the following conclusions:

1. In terms of round trip delays iBatis takes lesser time. The slighter increase in time in case of hibernate can be accounted to the time taken for automatically generating the queries and the complex caching algorithms used by hibernate.
2. In terms of flexibility iBatis has an upper hand over hibernate.
3. Considering the learning curve iBatis has a smaller curve since it is more similar to JDBC.
4. Programming using iBatis requires an SQL guru in the team but while using hibernate in-depth knowledge in SQL is not required.
5. Considering the features provided by both the tools, hibernate is much stronger since it supports lazy fetching and mapping associations.